\documentclass[letterpaper]{emulateapj}

\newcommand{\kms}{\,km\,s$^{-1}$}         
\newcommand{\smy}{\,M$_\odot$\,yr$^{-1}$}\newcommand{\hi}{\ion{H}{1}}
\newcommand{\tm}{\tablenotemark}       \newcommand{\tn}{\tablenotetext}
\newcommand{\hst}{\emph{HST}}    \newcommand{\sw}{\ion{S}{2}}
     
\newcommand{\siw}{\ion{Si}{2}}   \newcommand{\sit}{\ion{Si}{3}}
     \newcommand{\ha}{H$\alpha$}
\newcommand{\hw}{\ion{H}{2}}     \newcommand{\msun}{$M_\odot$} 
\newcommand{\oi}{\ion{O}{1}}     
\newcommand{\qa}{\object{PG2112+059}}
\newcommand{\qb}{\object{RXJ2043.1+0324}}
\newcommand{\qc}{\object{RXJ2139.7+0246}}

\begin{document}
\shorttitle{The Smith Cloud}
\shortauthors{Fox et al.}
\title{On the Metallicity and Origin of the Smith 
High-Velocity Cloud\footnotemark[1]}
\footnotetext[1]{Based on observations taken under program 13840 
  of the NASA/ESA Hubble Space Telescope, obtained at the 
  Space Telescope Science Institute, which is operated by the Association of 
  Universities for Research in Astronomy, Inc., under NASA contract 
  NAS 5-26555, and under program GBT09A\_17 of the Robert C. Byrd Green Bank
  Telescope (GBT) of the National Radio Astronomy Observatory,
  a facility of the National Science Foundation operated under a cooperative 
  agreement by Associated Universities, Inc.}
\author{Andrew J. Fox$^2$, Nicolas Lehner$^3$, Felix J. Lockman$^4$,  
Bart P. Wakker$^5$, Alex S. Hill$^6$, Fabian Heitsch$^7$, 
David V. Stark$^{7,8}$, Kathleen A. Barger$^9$, 
Kenneth R. Sembach$^2$, \& Mubdi Rahman$^{10}$} 
\affil{$^2$ Space Telescope Science Institute, 3700 San Martin Drive,
  Baltimore, MD 21218\\
$^3$ Department of Physics, University of Notre Dame, 225 Nieuwland 
Science Hall, Notre Dame, IN 46556\\
$^4$ National Radio Astronomy Observatory, P.O. Box 2, Rt. 28/92, 
Green Bank, WV 24944\\
$^5$ Department of Astronomy, University of
 Wisconsin--Madison, 475 North Charter St., Madison, WI 53706\\
$^6$ Department of Astronomy, Haverford College, 370 Lancaster Ave.,
Haverford, PA 19041\\
$^7$ Department of Physics and Astronomy, University of North Carolina 
at Chapel Hill,
120 E. Cameron Ave., Chapel Hill, NC 27599\\
$^8$ Kavli Institute for the Physics and Mathematics of the Universe,
5-1-5 Kashiwanoha, Kashiwa, 277-8583, Japan\\ 
$^9$ Department of Physics and Astronomy, Texas Christian University, 
TCU Box 298840, Fort Worth, TX 76129\\
$^{10}$ Department of Physics and Astronomy, Johns Hopkins University,
3400 North Charles St., Baltimore, MD 21218}

\email{afox@stsci.edu}

\begin{abstract} 
The Smith Cloud is a gaseous high-velocity cloud (HVC) in an advanced 
state of accretion, only 2.9\,kpc below the Galactic plane and due to 
impact the disk in $\approx$27\,Myr. 
It is unique among HVCs in having a known distance (12.4$\pm$1.3 kpc) and a 
well-constrained 3D velocity (296\kms), but its origin has
long remained a mystery.
Here we present the first absorption-line measurements of its metallicity, 
using \hst/COS UV spectra of three AGN lying behind the Cloud 
together with Green Bank Telescope 21\,cm spectra of the same directions.
Using Voigt-profile fitting of the \sw\ $\lambda\lambda$1250, 1253, 1259 
triplet together with ionization corrections derived
from photoionization modeling, 
we derive the sulfur abundance in each direction; a weighted average of 
the three measurements gives [S/H]=$-$0.28$\pm$0.14,
or 0.53$^{+0.21}_{-0.15}$ solar metallicity. 
The finding that the Smith Cloud is metal-enriched lends support to 
scenarios where it represents recycled Galactic material, 
rather than the remnant of a dwarf galaxy or accreting intergalactic gas.
The metallicity and trajectory of the Cloud are both indicative of
an origin in the outer disk.
However, its large mass and prograde kinematics remain to be fully explained.
If the cloud has accreted cooling gas from the corona
during its fountain trajectory, as predicted in recent theoretical work,
its current mass would be higher than its launch mass, alleviating 
the mass concern.
\end{abstract}

\section{Introduction}
The gaseous halo of the Milky Way is home to a diverse population 
of high-velocity clouds (HVCs)
that do not co-rotate with the underlying disk. HVCs trace a variety of 
processes including inflow, outflow, tidal stripping, and condensation 
of coronal material (Wakker \& van Woerden 1997). 
Once the neutral and ionized phases of HVCs are accounted for, they
represent a
global inflow rate onto the Galaxy of $\approx$0.4--1.4\smy\ 
(Shull et al. 2009, Lehner \& Howk 2011, 
Putman et al. 2012), similar to the Galactic star 
formation rate. HVCs therefore represent the Galactic fuel supply,
and understanding their physical and chemical properties allows us to explore 
the mechanism(s) by which star formation is sustained.

The Smith Cloud (SC; see Figure 1; Smith 1963), also known as the 
Galactic-Center-Positive (GCP) complex, is a large \
HVC\footnotemark[2] 
plunging toward the Galactic plane with a velocity $v_z$=73\kms, 
\footnotetext[2]{Parts of the SC are at intermediate velocities 
$v_{\rm LSR}\!<\!90$\kms, but it has traditionally been defined
as an HVC since its head is at $v_{\rm LSR}\approx100$\kms.}
an \hi\ mass of $\ga$10$^6$\msun\ 
(Lockman et al. 2008), and a similar \hw\ mass (Hill et al. 2009).
Its cometary morphology suggests that its direction
of motion lies along its major axis, and so 
its LSR velocity can be used to derive
a 3-D velocity of 296\kms\ (Lockman et al. 2008),
lower than the Galactic escape velocity.
The SC has a heliocentric distance of $d$=12.4$\pm$1.3\,kpc, 
based on three independent methods:
the detection of absorption lines in background stellar spectra
(Wakker et al. 2008), its kinematics (Lockman et al. 2008), 
and its H$\alpha$ emission (Putman et al. 2003).
It also has a magnetic field of $>$8$\mu$G detected via Faraday rotation 
(Hill et al. 2013) 
but no significant stellar population \citep{St15}.
Together these properties make the SC arguably 
the best characterized of all HVCs \citep[see][]{Lo15}.

\begin{figure}
\epsscale{1.25}\plotone{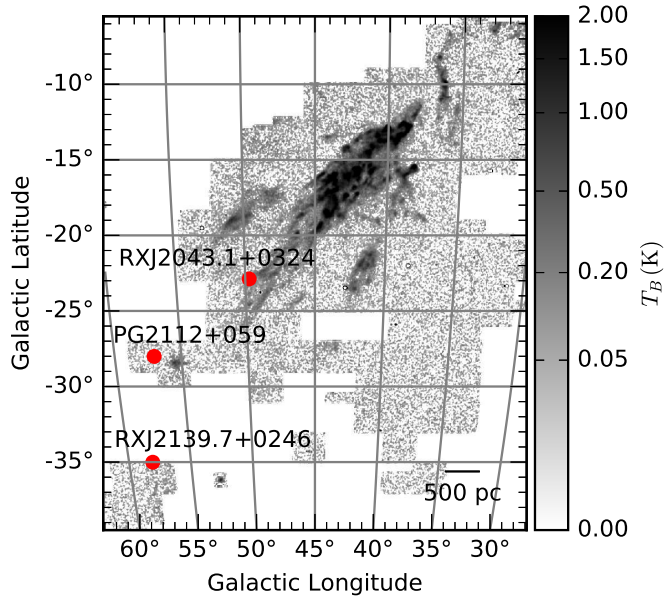} 
\caption{Map of the SC and its surroundings based on Hill et al. (2013).
The gray-scale shows the brightness temperature of 21\,cm H~I emission from
the Green Bank Telescope at a fixed velocity $v_{\rm GSR}$ = +247\kms\ in the 
Galactocentric standard of rest frame.
The three COS targets are shown with red circles.
All three probe the major axis of the SC, which is moving toward
the upper right in this image, and all three show detections of 21\,cm 
emission from the cloud or its wake.}
\end{figure}

Despite the good characterization of the SC's properties, its origin
remains unknown. It is unclear whether it represents Galactic material, 
expelled from the disk as part of a Galactic fountain (Bregman 1980), 
accreting extragalactic 
material, the gaseous remnant of a dwarf galaxy, such as the
Sagittarius dwarf (Bland-Hawthorn et al. 1998) or other satellite 
(Stark et al. 2015), or something else.
The key missing piece of evidence is the SC metallicity. 
Metallicity discriminates between recycled Galactic gas 
(enriched) and infalling extragalactic gas (unenriched), but since 
the SC lies close to the disk and so has relatively high foreground 
extinction, there have been (until now)
no AGN lying behind the SC observed in the UV. 
Thus there is no published absorption-line constraint on the SC metallicity, 
though Hill et al. (2009) measured the \ha\ and [\ion{N}{2}] emission from the 
SC using optical nebular lines and derived a nitrogen abundance of 
0.15--0.44 times solar. 

In this Letter we present absorption-line metallicity measurements
from three SC sightlines, toward the AGN \qa, \qb, and \qc. 
We describe the UV and radio observations in \S2, present the spectra and 
abundance calculations in \S3, and discuss the implications in \S4.
All velocities are presented in the LSR reference frame.

\section{Observations and Data Reduction}
The targets for this program were selected by searching for UV-bright AGN
behind or nearby the SC. This search yielded three targets (see Table 1).
All three probe the wake region of the SC, behind its direction of motion
(Figure 1), where there is extensive diffuse \hi\ both morphologically 
and kinematically associated with the cloud \citep{Lo08}.
The three AGN were observed in October 2014 with the 
Cosmic Origins Spectrograph \citep[COS;][]{Gr12} onboard \hst, under 
Program ID 13840 (PI Fox), with three orbits per target.
The spectra were taken with the G130M/1291 grating/central-wavelength 
combination, the primary science aperture, and all four FP-POS positions.
The {\tt x1d} files produced by the {\tt calcos} reduction pipeline
were aligned using customized reduction software, designed to ensure that 
commonly observed ISM lines are aligned in wavelength space (see Wakker 
et al. 2015 for more details). The data were binned by three pixels and 
continuum normalized for display.

\begin{deluxetable*}{lcccc ccccc}
\tablecaption{Properties of AGN behind the Smith Cloud}
\tabletypesize{\small}
\tabcolsep=2.0pt
\tablehead{Target & $l$\tm{a} & $b$\tm{a}  & $z_{\rm em}$\tm{b} & FUV\tm{c} &
$F_{1300}$\tm{d} & log\,$N$(H I)\tm{e} & $v_{\rm min}$\tm{f} & $v_{\rm max}$\tm{f} & $v_0$\tm{g}\\
& (\degr) & (\degr) & & & (flux units) & (cm$^{-2}$) & (km\,s$^{-1}$) & (km\,s$^{-1}$) & (km\,s$^{-1}$)}
\startdata
PG2112+059     & 57.04 & $-$28.01 & 0.457 & 17.05 & 0.75 & 18.72$\pm$0.06 & 40 &  75 & 42\\ 
RXJ2043.1+0324 & 49.72 & $-$22.88 & 0.271 & 17.29 & 0.55 & 18.84$\pm$0.05 & 40 & 110 & 79\\ 
RXJ2139.7+0246 & 58.09 & $-$35.01 & 0.260 & 16.79 & 0.79 & 19.41$\pm$0.02 & 40 & 100 & 55\\ 
\vspace{-3mm}
\enddata
\tn{a}{Galactic longitude and latitude.}
\tn{b}{Emission-line redshift of AGN.}
\tn{c}{GALEX FUV magnitude.}
\tn{d}{Spectroscopic  flux at 1300\,\AA\ in units of 10$^{-14}$
erg\,cm$^{-2}$\,s$^{-1}$\,\AA$^{-1}$.} 
\tn{e}{\hi\ column density in SC measured from GBT, integrated in range
$v_{\rm min}$ to $v_{\rm max}$, except for PG2112+059 value, determined by 
profile fitting.}
\tn{f}{Minimum and maximum LSR velocities of SC emission and absorption.}
\tn{g}{Central LSR velocity of SC 21\,cm emission.}
\end{deluxetable*}

\hi\ 21\,cm spectra were taken with the 100-meter Robert C. Byrd Green Bank
Telescope (GBT) at an angular resolution of $9\farcm1$, under 
project code GBT09A\_17.
Spectra were taken using in-band frequency switching, and
covered 700\kms\ centered at zero velocity LSR at a velocity 
resolution of 1.3\kms.
The data were calibrated and corrected for stray radiation
following the procedure described in Boothroyd et al (2011), 
and a low-order polynomial was fit to emission-free regions of the spectra 
to remove residual instrumental baseline structure.
The \hi\ column density in each direction
is calculated by integrating the 21\,cm brightness-temperature profiles 
(shown in Figure 2) between $v_{\rm min}$ and $v_{\rm max}$ by the 
standard relation

\begin{equation}
N({\rm H\,I})=1.823\times10^{18}{\rm cm}^{-2}\int^{v_{\rm max}}_{v_{\rm min}} T_B\,{\rm d}v.
\end{equation}

For the \qa\ direction, the SC component is blended on the blue (low-velocity)
side by Galactic emission, so in this case we determined the \hi\ 
column density by fitting a Gaussian component, with a velocity centroid 
matching that seen in \sw\ absorption.

\section{Results}

Among the many low-ionization lines in the COS G130M bandpass, 
\sw\ $\lambda\lambda$1250, 1253, 1259, and 
\oi\ $\lambda$1302 lines are the most useful for metallicity 
measurements, because sulfur and oxygen are relatively undepleted
onto dust and have relatively small ionization corrections.
Unfortunately, the \oi\ $\lambda$1302 lines from the SC are of very
limited use for metallicity measurements, because of strong saturation.
This leaves the \sw\ triplet as our best metallicity indicator.
\hst/COS absorption-line profiles showing \sw\ for the 
three sightlines under study are plotted in Figure 2, together
with Voigt-profile fits to the data. These were conducted
via simultaneous fits to unblended regions of the spectra,
taking account of the COS line spread function, 
and leaving all the initial inputs (velocity centroid, line width,
and column density) free to vary \citep[following][]{Le11}.
Clear \sw\ absorption from the SC is seen in each direction.

\begin{deluxetable*}{lcccc ccc}[!h]
\tablecaption{Smith Cloud Metallicity Measurements}
\tabletypesize{\small}
\tabcolsep=2.0pt
\tablehead{Target & log\,$N$(H~I) & log\,$N$(S~II)\tm{a} & [S~II/H~I]\tm{b} & log\,$\frac{N{\rm(Si\,III)}}{N{\rm(Si\,II)}}$\tm{c}    
& log\,$U$\tm{d} & IC(S)\tm{e} & [S/H]\tm{f}}
\startdata

PG2112+059     & 18.72$\pm$0.06 & 14.40$\pm$0.33 & +0.56$\pm$0.33  & ...\tm{g} & $\approx-3.0$ & $-$0.65$\pm$0.10 & $-$0.09$\pm$0.33$\pm$0.15\\ 
RXJ2043.1+0324\tm{h} & 18.84$\pm$0.05 & 14.38$\pm$0.13 & +0.42$\pm$0.13  & $\ga-0.89$ & $-$3.2$\pm$0.1 & $-$0.56$\pm$0.10 & $-$0.14$\pm$0.13$\pm$0.15\\   
RXJ2139.7+0246 & 19.41$\pm$0.02 & 14.17$\pm$0.20 & $-$0.36$\pm$0.20 & $\ga-0.99$ & $-$3.0$\pm$0.1 & $-$0.22$\pm$0.10 & $-$0.58$\pm$0.20$\pm$0.15\\ 
\vspace{-2mm}
\enddata
\tablecomments{We use the solar sulfur abundance (S/H)$_\odot$=$-$4.88 
from Asplund et al. (2009).} 
\tn{a}{Column density derived from Voigt-profile fitting (see Figure 2).} 
\tn{b}{\sw\ ion abundance, defined in Equation 2.}
\tn{c}{\sit/\siw\ column-density ratio, measured in the velocity interval 
$v_{\rm min}$ to $v_{\rm max}$.}
\tn{d}{Logarithm of best-fit ionization parameter, derived to match 
\sit/\siw\ ratio.}
\tn{e}{Ionization correction derived from {\it Cloudy} photoionization 
models (Figure 3).}
\tn{f}{S abundance corrected for ionization, [S/H]=[\sw/\hi]+IC(S). 
The first error is statistical. The second is systematic and accounts
for the beamsize mismatch between UV and radio observations, and 
uncertainties in the ICs.}
\tn{g}{Contamination prevents \siw/\sit\ ratio being measured; 
we adopt the same log\,$U$ value as derived for the RXJ2139.7+0246 sightline.}
\tn{h}{This sightline shows a $\approx$20\kms\ offset between the \sw\ and GBT
\hi\ centroids (see Figure 2).}
\end{deluxetable*}

\begin{figure*}
\epsscale{1.0}\plotone{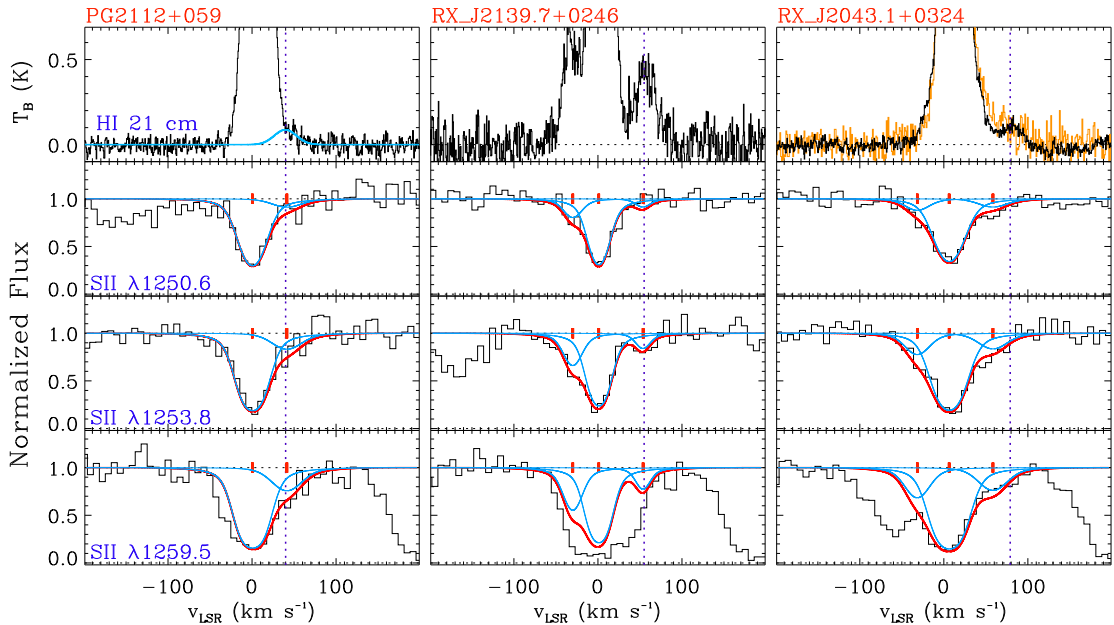}
\caption{\hst/COS S~II $\lambda\lambda$1250,1253,1259 absorption-line profiles
of the three AGN in our sample, together with GBT 21\,cm H I emission 
profiles (top panels).
In the case of RXJ2043.1+0324, the LAB 21\,cm data are shown in yellow.
The red solid line shows Voigt-profile fits to the COS data, which are binned
by three pixels for display. The fits are composed of two or three individual 
components, plotted as blue solid lines, with centroids denoted with tick marks.
The positive-velocity component
is due to the Smith Cloud; it covers the range $\approx$40--100\kms.
The dashed vertical line in each panel shows the central velocity of the SC 
in the GBT data.
Good alignment between the S~II absorption and GBT H~I emission is seen in 
the first two sightlines, but there is a $\approx$20\kms\ offset in the
third sightline. This is likely related to small-scale structure in the 
GBT beam, which is already suggested by the difference between the LAB and 
GBT profiles.}
\end{figure*}

An important issue in abundance determinations is velocity alignment.
The \sw\ and \hi\ lines should line up in velocity space
if they originate in the same cloud. For \qa, 
weak 21 cm emission from the SC is visible in the range $\sim$30--80\kms\
(see Figure 2), matching the range where absorption is seen in \sw;
the fitted SC \sw\ component has a centroid of 41.5$\pm$14.5\kms, 
with no evidence for misalignment from \hi.
For \qc, the 21 cm component centroid (55\kms)
and \sw\ component centroid (53.7$\pm$5.3\kms) are closely aligned.
However, for \qb, there is an $\approx$20\kms\ offset between the 
\sw\ component at 58.5$\pm$6.1\kms\ and the GBT \hi\ component 
centered at 79\kms.
To explore this, we downloaded the Leiden-Argentine-Bonn (LAB) 
21\,cm data of this sightline \citep{Ka05}, and found the 
SC emission in the LAB data was centered at 62\kms, 
within the 1$\sigma$ error of the \sw\ centroid. 
The GBT data (9.1\arcmin\ beam) 
have higher angular resolution than the LAB data (36\arcmin\ beam),
and have better sampling, but the fact that the GBT and LAB centroids 
disagree shows that small-scale structure (clumping)
exists in the neutral gas in this direction.
Indeed, we know there is a large gradient in $N$(\hi) in this
sightline since it passes just outside the main body of the SC (Figure 1).
This structure limits the accuracy of abundances derived by comparing 
pencil-beam metal columns with finite-beam \hi\ columns;
we add a systematic error of $\approx$0.10\,dex 
to the abundance calculations to account for this beamsize mismatch
\citep{Wa01}.

The sulfur abundance in the Smith Cloud is calculated from the observed 
\sw\ ion abundance using an ionization correction (IC). The ion abundance 
is determined observationally as:

\begin{equation}
[{\rm S~II/H~I}]=[{\rm log}\,N({\rm S~II})-{\rm log}N({\rm H~I})]-{\rm log}({\rm S/H})_\odot
\end{equation}
and the IC is defined such that
\begin{equation}
[{\rm S/H}]=[{\rm S~II/H~I}]+{\rm IC(S)}.
\end{equation}

The magnitude of the IC in a given direction depends on the \hi\ column density,
the ionization parameter $U\equiv n_{\gamma}/n_{\rm H}$
(the ratio of the ionizing photon density to the gas density),
and the shape and normalization of the incident ionizing radiation field.
We ran a grid of {\it Cloudy} photoionization models \citep{Fe13}
to investigate the magnitude of the ICs in the SC given a 3D model of the 
Galactic ionizing radiation field \citep{BM99, Fo05, Fo14}, as a function
of $N$(\hi) and $U$ (see Figure 3).
These models use the radiation field interpolated at the location where each
sightline intercepts the SC (given that we know $l$, $b$, and $d$). This field
has log\,$n_{\gamma}$=$-$4.70, corresponding to an ionizing flux 
$\Phi$=10$^{5.78}$\,photons\,cm$^{-2}$\,s$^{-1}$.
We constrain the value of $U$ by matching the
\sit/\siw\ column-density ratio in the SC,
log [$N$(\sit\ 1206)/$N$(\siw\ 1304)]$\ga-0.89$ in the \qb\ direction and
$\ga-0.99$ in the \qc\ direction, 
as measured from apparent optical depth (AOD) integrations \citep{SS91}
in the SC velocity interval. This process gives log\,$U$=$-$3.2$\pm$0.1
and $-$3.0$\pm$0.1, respectively, for these two sightlines
\citep[in line with values derived for other HVCs;][]{Co05, TS12, Fo14}.
The IC for sulfur is then calculated directly for each sightline 
(via Equation 3) using the model run at the appropriate $N$(H I) and $U$.
Any hidden saturation in \sit\ 1206 (or \siw\ 1304) would raise (or lower) the 
\sit/\siw\ ratio and push the solution for $U$ to slightly
higher (or lower) values. However, because the IC is insensitive to $U$, 
particularly at log\,$U\!\ga\!-3$, 
(see Figure 3, lower panel), this would not significantly change the IC.

\begin{figure}
\epsscale{1.2}\plotone{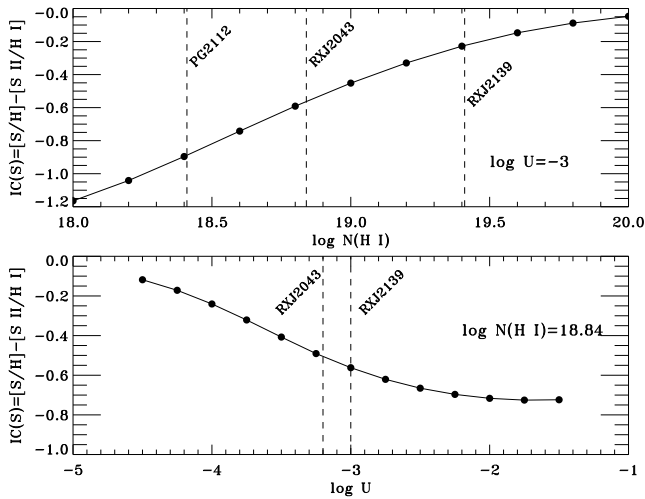}
\caption{The ionization correction IC(S) as a function of 
H I column density ({\bf top}) and ionization parameter ({\bf bottom}), 
used to convert the measured [S II/H I] ratios to [S/H] abundances. 
The ICs were calculated using {\it Cloudy} photoionization models 
and a Galactic ionizing radiation field calculated
at the position $l$=45\degr, $b$=$-$20\degr, $d$=12.5\,kpc,
which has an ionizing photon density 
log\,($n_\gamma$/cm$^{-3}$)=$-$4.70 \citep{Fo14}.
In the top panel, the ICs were calculated at 
log\,$U$=$-$3.0, the value required to match the observed 
Si III/Si II ratio in the Smith Cloud, and the observed H I column 
densities in the three sightlines are shown with vertical lines.
In the bottom panel, the ICs were calculated at a log $N$(H I)=18.84,
appropriate for the RXJ2043.1+0324 sightline.}
\end{figure}

The sulfur abundance measurements calculated using Equations 2 and 3
are summarized in Table 2.
Among our three SC sightlines, the \qc\ direction has a high enough 
\hi\ column density in the SC, log\,$N$(\hi)=19.41, that the 
ionization correction is relatively small ($-$0.22\,dex), giving
[S/H]=$-$0.56$\pm$0.20(stat)$\pm$0.15(syst), 
where the statistical uncertainty reflects the
measurement error on the \sw\ column density, and the 
systematic uncertainty derives from the UV-radio beamsize mismatch
and the uncertainties in the ionization correction.
Toward \qa\ and \qb, the ICs are larger ($-$0.65 and $-$0.56\,dex, 
respectively) because of lower \hi\ columns,
giving 
[S/H]=$-$0.09$\pm$0.33(stat)$\pm$0.15(syst) and 
[S/H]=$-$0.14$\pm$0.13(stat)$\pm$0.15(syst) in these two directions.
If there is no variation in the true metallicity across the cloud,
a weighted mean of the three measurements can be used, giving 
[S/H]=$-$0.28$\pm$0.14 (equivalent to 0.53$^{+0.21}_{-0.15}$ solar). 
However, the variation in abundance across the SC may be real. 
Indeed, the direction with the lowest derived abundance 
(toward \qc, where [S/H] is $\approx$0.5\,dex lower 
than in the other two sightlines) 
is the sightline furthest into the SC wake, where more metal mixing
with the surrounding gas is expected to occur \citep{Gr14}. 
Thus we see tentative evidence for an abundance gradient.

For comparison, the N abundance in the $\sim$10$^4$\,K ionized gas in 
the leading edge of the SC is 0.15--0.44 solar, as
derived from optical [\ion{N}{2}] emission lines and
assuming the ionized gas is photoionized
(although shocks may play a role there; Hill et al 2009, 2013). 
In a downstream portion of the cloud, more analogous to the sightlines 
probed in this work, the derived N abundance is 0.3--0.8 solar 
(Putman et al 2003, Hill et al 2013). 
These results suggest that the SC has a solar or sub-solar N/$\alpha$ ratio.

\section{Discussion and Summary}
The SC directly addresses a question of broad general interest:
how does gas get into galaxies? 
Since the SC is a coherent \hi\ cloud that has survived to its current
location without breaking apart, it clearly exemplifies one pathway to 
bring fuel into galaxies. The question is what is the origin of that pathway?

Our measurement of a high metallicity (0.53$^{+0.21}_{-0.15}$ solar) 
argues in favor of a Galactic origin for the SC,
such as gas on the returning leg of a fountain (Bregman 1980) 
launched into the halo at a different location in the disk (Sofue et al. 2004).
Given the Galactic radial chemical abundance gradient of
$-$0.06\,dex\,kpc$^{-1}$ \citep{HW99}, half-solar metallicity is
reached at a Galactocentric radius of 13 kpc, which is
exactly where the last SC disk passage occurred in the Lockman et al. (2008)
trajectory. {\it Therefore the metallicity and orbit of the Smith Cloud are 
both consistent with an origin in the outer disk.}

If the SC was accreting for the first time from the IGM,
or is the remnant ISM of a dwarf galaxy (Bland-Hawthorn et al. 1998), 
it would have a lower metallicity. The LMC has $\approx$0.50 solar 
metallicity, but is orders of magnitude more massive than the SC, 
and all other Local Group dwarfs have lower metallicity.
Stark et al. (2013) noted how the newly discovered star-forming galaxy 
Leo P has similar properties to the SC, and would also not show stars 
if at the SC's location. However, Leo P has a metallicity of 3\% solar 
\citep{Sk13}. Our SC metallicity conclusively rules out such a galaxy 
as the cloud's origin.

Our abundance calculations have not taken into account the depletion
of sulfur onto dust grains \citep{Je09}. However, any such depletion would 
\emph{raise} the total (gas+dust) inferred sulfur abundance, 
strengthening the conclusion that the SC is Galactic.
Furthermore, if any metal mixing has occurred in the Cloud's past,
so that it represents a diluted mixture of Galactic and extragalactic material,
the fact that the resulting metallicity is as high as $\approx$half solar
again strengthens the idea that Galactic (enriched) gas contributed to its 
origin (see Bland-Hawthorn et al. 2015 and Webster et al. 2015 for  
recent arguments concerning metal mixing).

It has long been a mystery how the SC survived to reach its current location,
because infalling HVCs are predicted and observed to be disrupted by their
interaction with the surrounding gaseous medium.
The lifetime against disruption depends on the clouds mass \citep{HP09}
and density contrast with the external medium \citep{Jo12}.
One possibility, explored by \citet{NB09} and \citet{GS15},
is that the SC might be a 
dark galaxy, bearing dark matter and gas but no stars. 
The dark matter would provide
the confinement that allowed the cloud to survive the 
$\approx$70 Myr since its last disk passage.
Our detection of metals argues against this idea, since the metals
require star formation to have occurred.

Another possibility to consider is whether the SC is a dark matter halo
that accreted sufficient disk material on a previous passage to explain its 
high metallicity. Using the disk column of \citet{Ni14} and the cloud's 
trajectory, the SC would travel through an ISM mass of 
$M_{\rm ISM}\!\approx\!2.3\!\times10^5 (R_{\rm SC}/100{\rm pc})^2$\msun\
during a single pass through the disk, which is $\approx$10\% of the SC mass
for a projected radius $R_{\rm SC}$=100\,pc.
However, this assumes that the cloud incorporates \emph{all} of the gas 
mass it travels through, which is unrealistic. Simulations show that the 
accretion efficiencies are $\approx$0.1--1\%, depending on the time and
velocity (F. Heitsch et al. 2016, in prep.), making it challenging for
this scenario to explain our metallicity observations.

Although our sulfur abundance is supportive of a Galactic origin for the SC,
there remain two main hurdles for this hypothesis: the cloud's mass
and kinematics. 
The mass problem is that the SC's high mass ($\approx$2$\times$10$^6$\msun)
is much larger than that of known extraplanar Galactic \hi\ clouds, 
such as superbubble ``caps'', and makes it unlikely that any 
star-formation process in the disk is energetic enough to
explain the SC \citep{Hi09}.
For example, the cap to the Ophiuchus superbubble 
contains only $\approx$3$\times$10$^4$\msun\ of \hi\ \citep{Pi07}.
However, recent theoretical work on Galactic fountains has shown they 
can sweep up and cool coronal gas \citep{Ma10, Ma13}.
In such models the mass of fountain HVCs grows with time
as they accrete cooled coronal material. 
In the hydrodynamical simulations of \citet{Fr15}, an HVC can 
increase its mass by a factor of three in 200 Myr.
The mass problem would be alleviated by such a mechanism, because
then the current SC mass would be higher than its launch mass and so
the required launch energy would be lower than previously thought. 

The second problem is the SC's kinematics. There is direct evidence
in the GBT \hi\ data that the Cloud has a
line-of-sight velocity at least 70\kms\ greater than the Galactic halo
material it is encountering \citep{Lo08}. 
The SC orbit is prograde and inclined at a shallow angle 
relative to the Milky Way disk ($\approx$30\degr). 
This indicates that the cloud is moving faster than Galactic rotation, 
and such super-rotation is not commonly observed: in other galaxies a 
\emph{lag} in rotation of gaseous material thrown up above the plane 
is typically seen \citep{Bo05, Sa08}.
A super-rotating extraplanar cloud would be unique.

In conclusion, our sulfur abundance of 0.53$^{+0.21}_{-0.15}$ solar for the SC 
provides an important new clue on its origin and supports a Galactic 
(as opposed to extragalactic) explanation, effectively ruling out dwarf-galaxy 
and dark-galaxy origins. However, the cloud's mass and kinematics
require it to be a highly unusual Galactic cloud.
This enigmatic object is still to be fully explained.

\vspace {0.3cm}
{\it Acknowledgments.}
The authors are grateful to Robin Shelton, 
Ken Croswell, and the referee for valuable comments.
Support for program 13840 was provided by NASA 
through grants from the Space Telescope Science Institute, which is 
operated by the Association of Universities for Research 
in Astronomy, Inc., under NASA contract NAS~5-26555.

\end{document}